\documentclass{PoS}
\usepackage[pdftex]{graphicx}                                                     
\usepackage{amsmath}
\usepackage{amssymb}
\usepackage{latexsym}
\usepackage{amsfonts}
\usepackage{longtable}
\usepackage{epstopdf}
\usepackage{url}
\usepackage{placeins}
\usepackage{tikz}
\usepackage{color}
\usetikzlibrary{positioning}

\title{Positive parity $D_s$ mesons}

\ShortTitle{Positive parity $D_s$ mesons}

\author{\speaker{Luka Leskovec}\\
        Jozef Stefan Institute, Ljubljana, Slovenia\\
        Department of Physics, University of Arizona, Tucson, Arizona, USA\\
        E-mail: \email{leskovec@email.arizona.edu}}

\author{C.B. Lang\\
        Institut f\"ur Physik, University of Graz, Graz, Austira\\
        E-mail: \email{christian.lang@uni-graz.at}}

\author{Daniel Mohler\\
        Fermi National Accelerator Laboratory, Batavia, Illinois, USA\\
        Helmholtz-Institut Mainz, Johannes Gutenberg-Universit\"at Mainz, Germany\\
        E-mail: \email{dmohler@fnal.gov}}

\author{Sasa Prelovsek\\
        Jozef Stefan Institute, Ljubljana, Slovenia\\
        Faculty of Mathematics and Physics, University of Ljubljana, Slovenia\\
        E-mail: \email{sasa.prelovsek@ijs.si}}

\author{R.M. Woloshyn\\
        TRIUMF, Vancouver, Canada\\
        E-mail: \email{rwww@triumf.ca}}

\abstract{We study the positive parity charmed strange mesons using lattice QCD,
the only reliable {\it ab initio} method to study QCD at low energies. Especially the 
experimentally observed $D_{s0}^*(2317)$ and $D_{s1}(2460)$ have challenged theory
for quite some time. The dynamical lattice QCD simulations are performed at two distinct 
pion masses, $m_{\pi} = 266$ MeV and $156$ MeV, using both $\bar{c}s$ as well as 
$DK$ and $D^*K$ scattering operators in the construction of the correlation matrix 
in order to take into the account threshold effects. While the $J^P = 0^+$ channel benefited 
most from the inclusion of scattering operators, it was also crucial for the case
of the $D_{s1}(2460)$. Using the L\"uscher method, which relates the finite volume spectrum
to the infinite volume scattering matrix, we were able to determine the near threshold 
behavior of the scattering amplitude. From it we extracted the binding momenta and the 
masses of the below-threshold bound states $D_{s0}^*(2317)$ and $D_{s1}(2460)$ by 
determining the pole positions of the scattering amplitudes. Our results compare well with 
experiment, resolving a long standing discrepancy between theory and experiment.}

\FullConference{International Symposium on Lepton Photon Interactions at High Energies\\
		17-22 August 2015\\
		University of Ljubljana, Slovenia}

\begin{document}

\section{Introduction}

The positive parity $D_s$ mesons, especially the $D_{s0}(2317)$ and $D_{s1}(2460)$, have until recently presented an unsolved puzzle in theoretical physics. Experimentally their masses lie below the $DK$ and $D^*K$ thresholds respectively \cite{Agashe:2014kda}, however neither quark models nor lattice QCD studies were able to reproduce this phenomenon uncontroversially (see Refs. \cite{Mohler:2013rwa,Lang:2014yfa} and references therein). Early quenched lattice studies, that ignored sea quark contributions, only took into account $c\bar{s}$ operators and found results consistent with the quark model expectations: $D_{s0}(2317)$ and $D_{s1}(2460)$ masses appeared above the $DK$ and $D^*K$ thresholds respectively \cite{Lewis:2000sv}. Recent dynamical studies investigated whether the issue might lie in the sea quark contributions (or lack thereof), however when the pion and kaon masses in dynamical lattice QCD studies were close to their physical values, the low lying positive parity $D_s$ mesons again had masses above their respective thresholds \cite{Mohler:2011ke}.

In this work we used lattice QCD as an {\it ab initio} method to study hadronic physics, to investigate the masses of positive parity $D_s$ mesons: $D_{s0}(2317)$ and $D_{s1}(2460)$ with $J^P=0^+$ and $J^P=1^+$ respectively. Lattice QCD is the theory of the strong interaction formulated in discrete Euclidean space-time and its main advantage is that the correlation functions of hadronic operators are evaluated in terms of fundamental QCD degrees of freedom: quarks and gluons. The calculations are performed in a finite sized box with periodic boundary conditions, rendering the discrete spectrum of correlation functions.

Lattice QCD studies of the positive parity mesons thus far have only included $c\bar{s}$ operators\footnote{Often referred to as the single hadron approach.} in the construction of the correlation functions. In our study of the positive parity $D_s$ mesons we include also the $DK$ and $D^*K$ scattering operators as well as $c\bar{s}$ operators to investigate whether the coupling of $D_{s0}^*(2317)$ to the $DK$ threshold and the coupling of $D_{s1}(2460)$ to the $D^* K$ threshold could affect their masses.

\section{Details of the lattice setup}

The simulation was performed on two ensembles of dynamic gauge field configurations \cite{Hasenfratz:2008ce,Hasenfratz:2008fg,Aoki:2008sm} and some details are presented in Table \ref{tab:gauge}.
\begin{table}[htb]
\begin{center}
    \begin{tabular}{c|l|l|l}
        quantity & Ensemble (1)\cite{Hasenfratz:2008ce,Hasenfratz:2008fg} & Ensemble (2)\cite{Aoki:2008sm} & experiment \cite{Agashe:2014kda} \cr
        \hline
        $N_L^3\times N_T$ ($N_f$) & $16^3\times32$ (2) & $32^3\times64$ (2+1) & \cr
        $L$[fm]     & 1.98(2)    & 2.90(4)    &  \cr
        \hline
        $m_\pi$[MeV]  & 266(3) & 156(7) &  139.57018(35) \cr
        $m_K$[MeV]    & 552(1) & 504(1) &   493.677(16)  \cr
    \end{tabular}
\end{center}
\caption{Details of gauge configurations used in our study of the positive parity $D_s$ mesons. $N_L$ and $N_T$ denote the number of spatial and temporal points on the lattice, while $L$ lists the spatial length of a single side of the box. $\pi$ and $K$ masses for both ensembles are compared to their experimental counterparts.}
\label{tab:gauge}
\end{table}

Due to the large charm quark mass we treat it differently than the light and strange quarks. To minimize the discretization errors associated with the large charm quark mass we use the Fermilab method \cite{ElKhadra:1996mp}, where the discretization effects are minimal in the difference between the energy of a given state $E$ and the $D_s$ spin averaged mass: $E - \overline{m}_{D_s}$, where 
\begin{align}
\overline{m}_{D_s}=\tfrac{1}{4}(m_{D_s}+3m_{D_s^*}).
\end{align}

\section{Spectrum}

The energy spectrum of the positive parity $D_s$ mesons in our calculation is comprised of discrete energy levels related to multihadron ($DK$ or $D^*K$ scattering) states and additional resonances and bound states. The spectrum is discrete due to periodic boundary conditions in space, where the expected noninteracting energies of two particle states are:
\begin{align}
  E=E_{D,D^*}(\vec{p}_1) + E_{K}(\vec{p}_2), \text{ where } |\vec{p}_i|=\vec{n}_i \frac{2\pi}{L} \text{ and } \vec{n}_i \in Z^3.
\end{align}
The correlation matrices $C_{ij}(t)$ for both channels built from operators $O_i$ listed in \cite{Mohler:2013rwa,Lang:2014yfa} with desired quantum numbers and symmetries:
\begin{align}
 C_{ij}(t)= \langle 0 | O_i(t_{source}+t)| O_j^{\dagger}(t_{source})|0\rangle = \sum_n Z_i^n Z_j^{*n} e^{-E_n t}
\end{align}
are evaluated using the distillation \cite{Peardon:2009gh} and stochastic distillation methods \cite{Morningstar:2011ka}. The spectrum ${E_n}$ of discrete scattering states and additional resonances/bound states is extracted from the correlation matrix by solving the generalized eigenvalue problem:
\begin{align}
 C(t)\vec{v}_{n}(t) = \lambda_n(t)C(t_0)\vec{v}_{n}(t),
\end{align}
where $\lambda_n(t) \propto e^{- E_n t}$ for $t \to \infty$.

\begin{figure}[htb]
\begin{center}
    \includegraphics[width=0.48\textwidth]{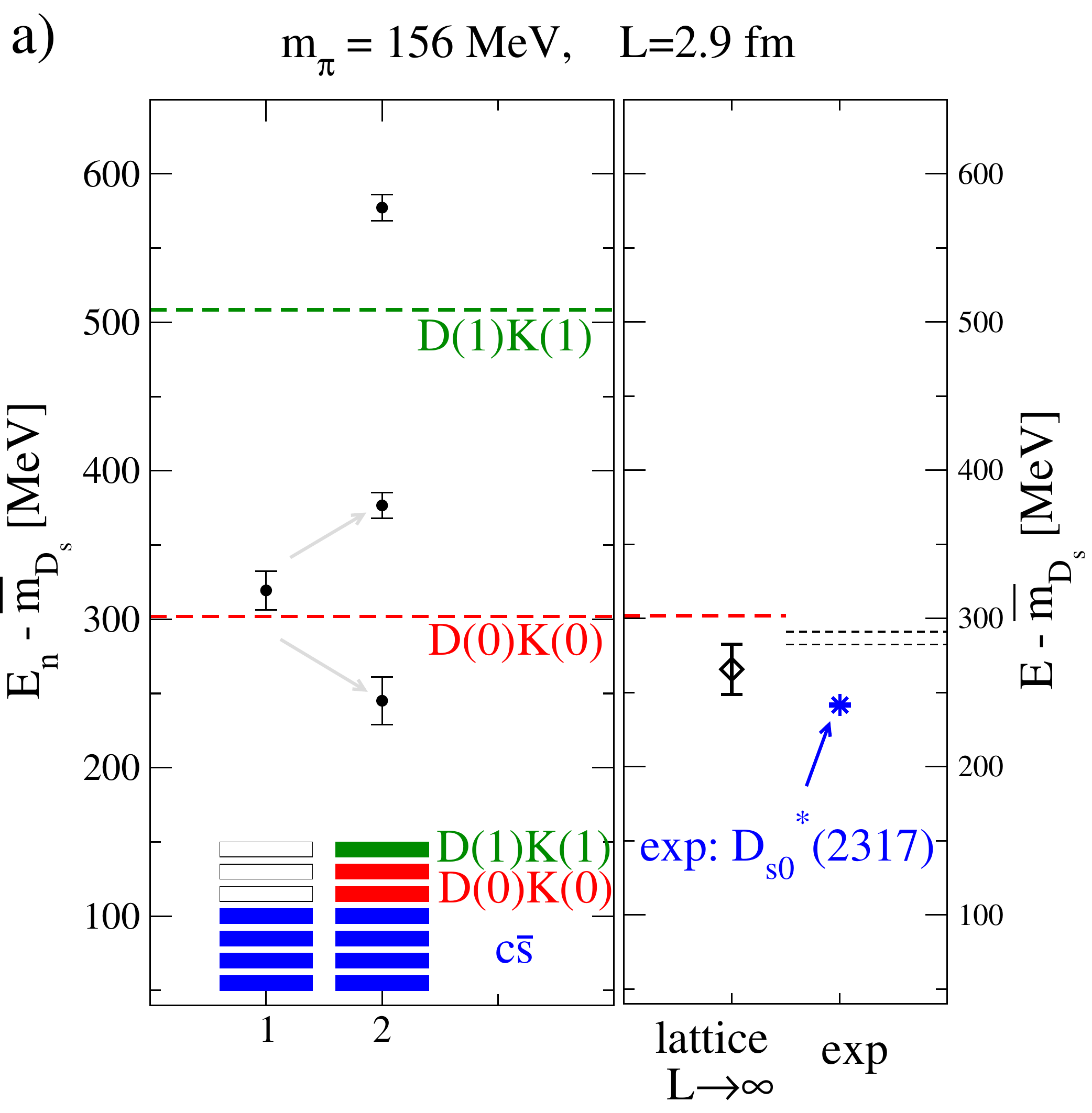}
    \includegraphics[width=0.5\textwidth]{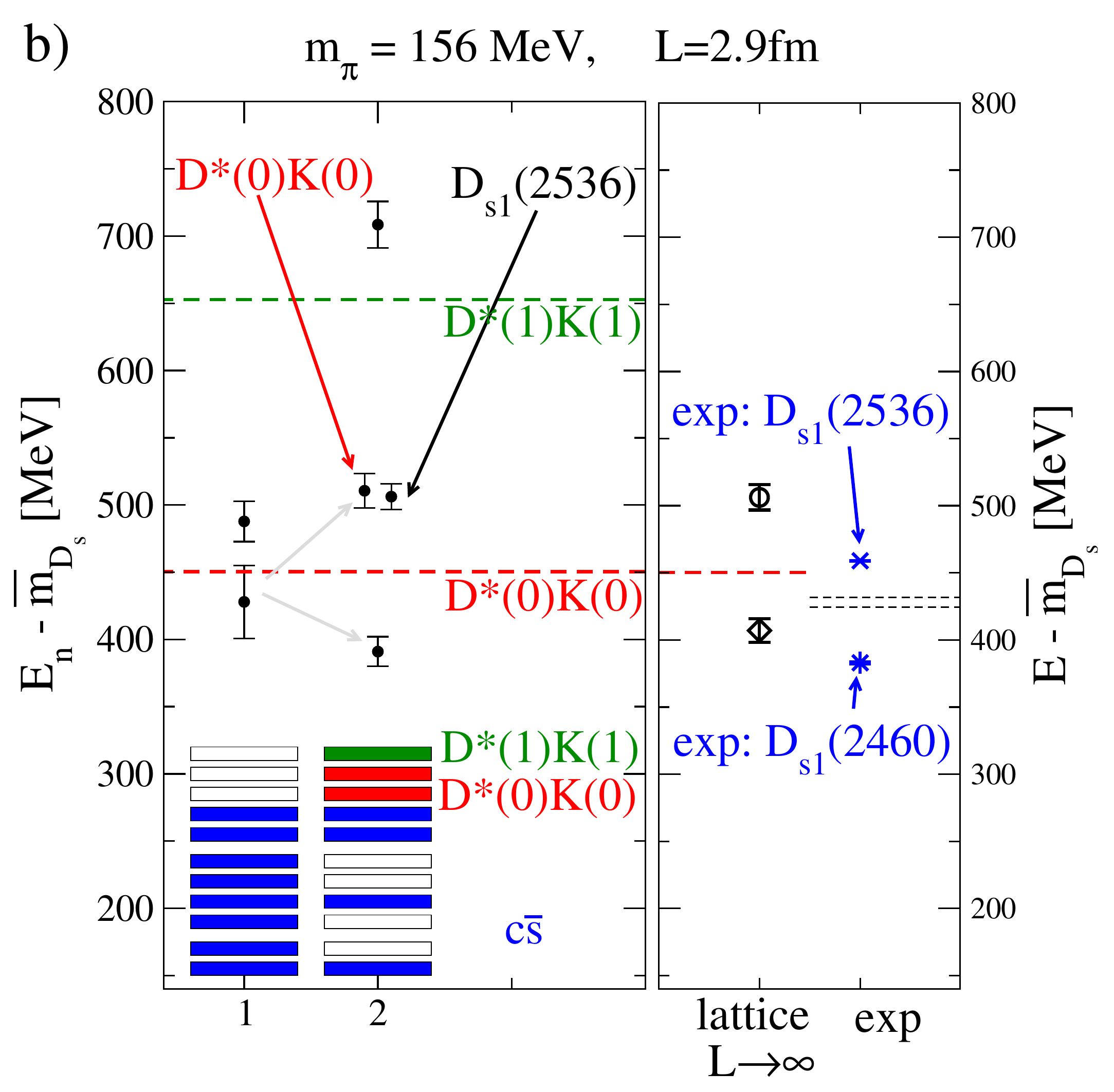}
\end{center}
\caption{The spectrum for the $J^P=0^+$ (left) and $J^P=1^+$ (right) $D_s$ meson channel on Ensemble (2).}
\label{fig:DS_spec}
\end{figure}

The $J^P=0^+$ and $J^P=1^+$ spectrum of the $D_s$ mesons is presented in Fig. \ref{fig:DS_spec}. In the $J^P=0^+$ channel we first calculated the correlation matrix with only $c\bar{s}$ operators. The resulting spectrum is shown in the left panel of Fig. \ref{fig:DS_spec}a. The energy level attributed to the $D_{s0}^*(2317)$ appears to be above the $DK$ threshold indicated by the dashed red line, fully consistent with previous studies. In the middle panel of Fig. \ref{fig:DS_spec}a we included the $DK$ scattering operators in the correlation matrix. The ground state energy is split into two energy levels, one below threshold and one above. A similar scenario occurred in the $J^P=1^+$ channel shown in Fig. \ref{fig:DS_spec}b, where the ground state with $c\bar{s}$ operators only appeared to be consistent with the the threshold energy (left panel). The ground state decoupled into two distinct states with the addition of $D^*K$ operators (middle panel).

To reliably disentangle the nature of the two states in both the $J^P=0^+$ and $J^P=1^+$ channels, we used the L\"uscher analysis \cite{Luscher:1990ux}, which relates the finite volume spectrum to the infinite volume scattering matrix. This allowed us to determine the $s$-wave scattering lengths and effective ranges for both channels, which are presented in Table \ref{tab:DS}.

\begin{table}[htb]
\begin{center}
\footnotesize
\begin{tabular}{c|c|c|c|c|c|c}
    quantity & $a_{0}^{0^+}$ & $r_{0}^{0^+}$ & $a_{0}^{1^+}$ & $r_{0}^{1^+}$ & $m_D + m_K -$                      & $m_{D^*} + m_K -$ \cr
                & [fm]             & [fm]             & [fm]             & [fm] & $m_{D_{s0}^*(2317)}^{BS}$[MeV] & $m_{D_{s1}(2460)}^{BS}$[MeV] \cr
             \hline
    Ensemble (1) & -0.756(25) & -0.056(31) & -0.665(25) & -0.106(37) & 78.9(5.4) & 93.2(4.7) \cr
    Ensemble (2) & -1.33(20)  & 0.27(17)   & -1.11(11)  & 0.10(10) & 36.6(16.6) & 44.2(9.9) \cr
    experiment   &   /  &   /  & / & / & 45.1 & 44.7 \cr
\end{tabular}
\normalsize
\caption{Scattering lengths, effective ranges and the distance to the $DK$ or $D^*K$ threshold for the $J^P=0^+$ and $J^P=1^+$ $D_s$ channels on Ensemble (1) and (2). For reference we also compare with experimental data from \cite{Agashe:2014kda}.}
\label{tab:DS}
\end{center}
\end{table}

The negative scattering lengths presented in Table \ref{tab:DS} demonstrate, that the $D_{s0}^*(2317)$ and $D_{s1}(2460)$ are indeed below threshold states. By plugging the effective range expansion in to the bound state pole condition $\cot{\delta}=i$, we find the masses of the two positive parity $D_s$ mesons. Since the quark masses are unphysical, we present the mass difference between the bound states and their respective thresholds in Table \ref{tab:DS}. The discrepancy between the mass differences on Ensembles (1) and (2) in Table  \ref{tab:DS} is due to the unphysical pion mass of Ensemble (1), which is $266$ MeV. The pion mass of Ensemble (2) is much closer to physical, $156$ MeV, and is thus in better agreement with experiment.

\section{Conclusions}

The mesons $D_{s0}^*(2317)$ and $D_{s1}(2460)$ lie near $DK$ and $D^*K$ threshold in experiment, and reliable determination of their masses in lattice QCD necessitates that the effects of threshold are taken into account. The energy levels assigned to the positive parity $D_s$ mesons in the single hadron approach were compatible with the $DK$ and $D^*K$ threshold energies. With the inclusion of multihadron operators, the ground states decoupled into two states mainly attributed to the s$D^{(*)}K$ and the other to the physical bound state. The analytical continuation of the scattering amplitude determined with the L\"uscher method established the existence of poles below threshold related to $D_{s0}^*(2317)$ and $D_{s1}(2460)$. The resulting masses were found compatible with the experimental data for the first time and thus resolving a long standing puzzle in hadronic physics.


\end{document}